\documentclass[twocolumn,pre]{revtex4}

\usepackage{dcolumn}
\usepackage{amsmath}

\usepackage{graphicx}
\usepackage{bbm}

\newcommand{\etal}{{\it{}et~al.}}
\newcommand{\defn}{\textit}

\newcommand{\mat}{\mathbf}

\begin{document}

\title{Network structure from rich but noisy data}

\author{M. E. J. Newman}
\affiliation{Department of Physics and Center for the Study of Complex Systems,
University of Michigan, Ann Arbor, MI 48109, USA}

\begin{abstract}
Driven by growing interest in the sciences, industry, and among the broader public, a large number of empirical studies have been conducted in recent years of the structure of networks ranging from the internet and the world wide web to biological networks and social networks.  The data produced by these experiments are often rich and multimodal, yet at the same time they may contain substantial measurement error.  In practice, this means that the true network structure can differ greatly from naive estimates made from the raw data, and hence that conclusions drawn from those naive estimates may be significantly in error.  In this paper we describe a technique that circumvents this problem and allows us to make optimal estimates of the true structure of networks in the presence of both richly textured data and significant measurement uncertainty.  We give example applications to two different social networks, one derived from face-to-face interactions and one from self-reported friendships.
\end{abstract}

\maketitle 

Most empirical studies of networks take a ``naive'' view of structural data, meaning that one assumes that the data \emph{are} the network.  For instance, in a study of a protein--protein interaction network~\cite{Uetz00,Ito01,Giot03} one might compile a list of known protein interactions and represent them as a network of protein nodes joined by interaction edges.  But this network represents the pattern of \emph{measured} interactions not the pattern of actual interactions.  The two could, and probably do, differ substantially, because of both error in the measurements and missing data~\cite{vonMering05,Krogan06,WPVS09}.  As another example, in studies of friendship networks~\cite{RH61,Resnick97} one commonly assembles a network simply by asking people who their friends are.  Different people, however, may use different standards for what constitutes a friendship or recall relationships differently, so that two edges in the network may represent quite different types of relationship~\cite{KB76,BK77,Marsden90,Butts03}.  If, as one commonly does, one nonetheless analyzes the network as if all edges were equivalent, then one's results and conclusions may be unreliable.

At the same time, many studies return data much richer than just a simple measurement of connections.  Protein--protein interaction networks, for example, are commonly assembled from the results of many complementary experiments involving a variety of laboratory techniques, further enriched by knowledge of protein function, genetics, or other features.  Friendship networks can likewise be probed in different ways, using surveys, online data, observations of face-to-face interactions, and others, possibly enhanced with metadata on participant location, occupation, age, and many other characteristics.  Taken together these many types of data may be able to give a more accurate and nuanced picture of network structure than any single one can alone.

Generically, the question we want to answer is this: given the results of a set of measurements performed on a system of interest, what is our best estimate of the structure of the underlying network?  The data could take any form.  They could be rich, hierarchical, multilevel, and multimodal, but they may also be unreliable and error prone.  Some of the data may have no bearing at all on the network structure.  Others maybe related only obliquely to it.  And we may not know in advance which data are relevant and which are not, or how accurate any of the measurements are.  Remarkably, under these seemingly daunting circumstances we can nonetheless make progress.

Suppose that we are interested in the structure of a certain $n$-node network and for the moment let us concentrate on the commonest case of an unweighted undirected network.  (We consider generalizations to weighted and directed data below, and in the Supplementary Materials.)  Let us denote the true structure of the network---which we do not know---by an $n\times n$ symmetric adjacency matrix~$\mat{A}$, having elements $A_{ij}=1$ if nodes $i$ and~$j$ are connected by an edge and 0 otherwise.  This structure, commonly called the \defn{ground truth}, is the thing we are trying to estimate.

We now make a set of measurements of the system, measurements that can take many forms as discussed above, perhaps including direct measurements of network structure but also potentially including indirect measurements, metadata, or ``red herrings'' that have nothing to do with the network at all.  The network structure and the data are related to one another by a \defn{data model}, expressed in the form of a probability function $P(\mbox{data}|\mat{A},\theta)$ that specifies the probability of making the particular set of measurements we did, given the ground-truth network~$\mat{A}$ plus, optionally, some additional model parameters, which we collectively denote by~$\theta$.  In general, we do not know the form of this probability distribution---in most cases it will be a complicated function---but the option to include parameters~$\theta$ allows us to specify a family of functions that encompass a broad spectrum of possibilities.  Our goal will be, given such a family, first to determine the values of the parameters, which fixes the relationship between the network structure and the data, and then, given those values, to estimate the network structure itself.

To achieve these goals we use the method of maximum likelihood.  Applying Bayes' rule, we write
\begin{equation}
P(\mat{A},\theta|\mbox{data}) = {P(\mbox{data}|\mat{A},\theta) P(\mat{A}) P(\theta)\over P(\mbox{data})}.
\label{eq:bayes}
\end{equation}
Summing over all possible network structures~$\mat{A}$ we get $P(\theta|\mbox{data}) = \sum_\mat{A} P(\mat{A},\theta|\mbox{data})$, which we then maximize to find the most probable value of the parameters~$\theta$ given the observed data.  In fact, for convenience, we maximize not $P(\theta|\mbox{data})$ but its logarithm, whose maximum falls in the same place.  Employing the well-known Jensen inequality $\log \sum_i x_i \ge \sum_i q_i \log(x_i/q_i)$, we can write
\begin{align}
\log P(\theta|\mbox{data}) &= \log \sum_{\mat{A}} P(\mat{A},\theta|\mbox{data})
  \nonumber\\
  &\ge \sum_{\mat{A}} q(\mat{A}) \log {P(\mat{A},\theta|\mbox{data})\over q(\mat{A})},
\label{eq:jensen}
\end{align}
where $q(\mat{A})$ is any probability distribution over networks~$\mat{A}$ satisfying $\sum_{\mat{A}} q(\mat{A}) = 1$.  It is trivially the case that exact equality between left- and right-hand sides of Eq.~\eqref{eq:jensen} is achieved when
\begin{equation}
q(\mat{A}) = {P(\mat{A},\theta|\mbox{data})\over \sum_{\mat{A}} P(\mat{A},\theta|\mbox{data})},
\label{eq:estep}
\end{equation}
and hence this choice maximizes the right-hand side with respect to~$q$.  A further maximization with respect to~$\theta$ will then give us the maximum-likelihood value we seek.  To put that another way, a double maximization of the right-hand side of~\eqref{eq:jensen} with respect to both $q$ and $\theta$ will give us our answer for~$\theta$.  This can be easily carried out by maximizing first with respect to~$q(\mat{A})$ using Eq.~\eqref{eq:estep} and then with respect to~$\theta$, repeating until the result converges.  Differentiating~\eqref{eq:jensen} while holding $q(\mat{A})$ constant, we find the maximum with respect to~$\theta$ to be the solution of
\begin{equation}
\sum_{\mat{A}} q(\mat{A}) \nabla_\theta \log P(\mat{A},\theta|\mbox{data}) = 0.
\label{eq:mstep}
\end{equation}

Our calculation consists of iterating Eqs.~\eqref{eq:estep} and~\eqref{eq:mstep} from random initial values to convergence.  The final result is a value for the parameters~$\theta$, which we can then use to estimate the ground-truth network.  In fact, however, it turns out that this last step is unnecessary: the calculations we have already performed give us the ground-truth network structure as a by-product, indeed they give us the entire posterior probability distribution over structures, since from Eq.~\eqref{eq:estep} the quantity~$q(\mat{A})$ is none other than $q(\mat{A}) = P(\mat{A},\theta|\mbox{data})/P(\theta|\mbox{data}) = P(\mat{A}|\mbox{data},\theta)$.  In other words it is precisely the probability of the network having true structure~$\mat{A}$ given the observed data and our estimate of the parameters~$\theta$.

The method derived here is an example of an expectation--maximization or EM algorithm~\cite{DLR77,MK08}.  As described the method is a general one that can be used with many different networks and data models.  Let us see how it is applied in practice.

Our first example application is to a social network of US university students.  The data come from the ``reality mining'' study of Eagle and Pentland~\cite{EP06}, which aimed to establish the real-world social network of a set of individuals by measuring their physical proximity over time.  The 96 students participating in the study were given mobile phones that used special software to record when they were in proximity with one another.  The resulting record of pairwise proximity measurements is both richer and poorer than a direct network measurement, in exactly the manner considered in this paper.  It is richer in the sense that interactions between individuals may be measured repeatedly and not just once, but poorer in the sense that proximity is an error-prone indicator of actual interaction---two individuals may find themselves coincidentally in proximity, as they pass on the street say, without being acquainted or having any social interaction.

We take as our data set the measurements made during the reality mining study for eight consecutive Wednesdays in March and April of 2005.  (We choose weekly observations to remove weekly periodic effects, and March and April because they fall during the university term.)  This gives us eight sets of observations, one for each day, in which an observed edge means that two individuals were in physical proximity at some time during that day.

The data model we adopt for these data is a particularly simple one, in which the edge measurements---the observations of proximity---are assumed to be independent identically distributed random variables, conditioned on the ground truth~$A_{ij}$.  That is, the probability of observing an edge between nodes~$i$ and~$j$ depends only on the matrix element~$A_{ij}$ and in the same way for all~$i,j$.  This dependence can be parametrized by two quantities: the \defn{true-positive rate}~$\alpha$ (also called the \defn{sensitivity} or \defn{recall}), which is the probability of observing an edge where one truly exists, and the \defn{false-positive rate}~$\beta$, the probability of observing an edge where none exists.  In addition, we will assume a uniform prior probability~$\rho$ of the existence of an edge in any position, so that our model is parametrized by three parameters $\alpha$, $\beta$, and~$\rho$.

\begin{figure*}
\begin{center}
\includegraphics[width=15cm]{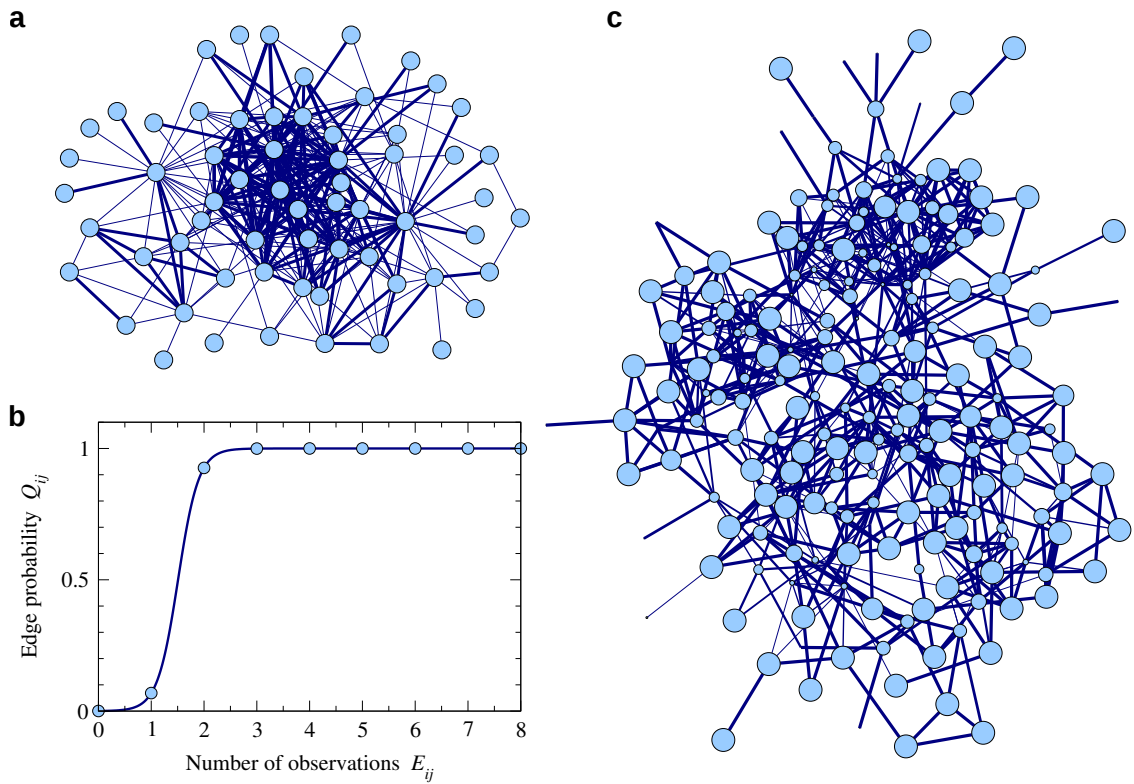}
\end{center}
\caption{(a)~Inferred ground-truth network for the reality mining data set.  Edge widths indicate the inferred probabilities~$Q_{ij}$.  Edges that are never observed are omitted, as are singleton nodes with no observed edges.  The figure reveals a dense core of about twenty nodes that are with high probability connected to one another and a sparser periphery of nodes for whom the surety of connection is much lower.  The thickest edges shown have $Q_{ij}>0.999$, while the thinnest have $Q_{ij}<0.1$.  (b)~Inferred edge probability as a function of the number of observations~$E_{ij}$ for the reality mining data set, showing a relatively sharp transition between $E_{ij}=1$ and $E_{ij}=2$.  (c)~Inferred network for the AddHealth friendship network.  Edge widths again indicate inferred probabilities, while node diameters are proportional to the so-called precision~$\rho\alpha_i/[\rho\alpha_i+(1-\rho)\beta_i]$, which is the estimated fraction of reported friendships that actually exist.  Some nodes are invisible because they are unreliable---their precision is very small---though these nodes may nonetheless have edges if another (reliable) node reports a connection.  Unobserved edges and singleton nodes are again omitted from the figure.}
\label{fig:results}
\end{figure*}

If for each node pair~$i,j$ we make~$N$ measurements and observe an edge to be present in~$E_{ij}$ of them then, as shown in the Methods, our EM equations give the following expressions for the three parameters:
\begin{equation}
\alpha = {\sum_{i<j} E_{ij} Q_{ij}\over N\sum_{i<j} Q_{ij}}, \qquad
\beta = {\sum_{i<j} E_{ij} (1-Q_{ij})\over N\sum_{i<j} (1-Q_{ij})},
\label{eq:params1}
\end{equation}
\begin{equation}
\rho = {1\over{n\choose2}} \sum_{i<j} Q_{ij},
\label{eq:params2}
\end{equation}
where $Q_{ij}$ is the posterior probability that there is an edge between~$i$ and~$j$, which is given by
\begin{equation}
Q_{ij} = {\rho \alpha^{E_{ij}} (1-\alpha)^{N-E_{ij}}\over
          \rho \alpha^{E_{ij}} (1-\alpha)^{N-E_{ij}} +
          (1-\rho) \beta^{E_{ij}} (1-\beta)^{N-E_{ij}}}.
\label{eq:qij}
\end{equation}
The full calculation involves iterating Eqs.~\eqref{eq:params1}, \eqref{eq:params2}, and~\eqref{eq:qij} until convergence is reached, and the results tell us the values of the three parameters~$\alpha$, $\beta$, and~$\rho$, as well as the entire posterior probability distribution over possible ground-truth networks, which is given by $P(\mat{A}|\mbox{data},\theta) = \prod_{i>j} Q_{ij}^{A_{ij}} (1-Q_{ij})^{1-A_{ij}}$.  The posterior distribution allows us to compute the distribution of any other network quantity we might be interested in---degrees, correlations, clustering coefficients, and so forth.  (See Section~\ref{sec:properties} in the Supplementary Materials.)

Applying Eqs.~\eqref{eq:params1}, \eqref{eq:params2}, and~\eqref{eq:qij} to the reality mining data, the algorithm converges rapidly and reliably to parameter values~$\alpha=0.4242$, $\beta=0.0043$, and $\rho=0.0335$.  The small value of~$\beta$ tells us that there are very few false positives: an edge is observed where none exists less than 1\% of the time.  On the other hand, even if the false-positive rate is low, the probability of being wrong when one does observe an edge can still be high.  This probability, called the \defn{false discovery rate}, is given by $(1-\rho)\beta/[\rho\alpha+(1-\rho)\beta]$ and is equal to~$0.2270$ in the present case, meaning that more than one in every five observed edges is in error.  Moreover, the relatively small value of~$\alpha$ implies that there are also a large number of false negatives: around 58\% of pairs of individuals who are in fact connected in the underlying network are not observed in proximity on any one day.  This is understandable.  Most people do not see all of their acquaintances every day.

Figure~\ref{fig:results}a shows the inferred ground-truth network, with edge thicknesses varying to indicate the probability~$Q_{ij}$ of individual edges.  In Fig.~\ref{fig:results}b we show the relationship between the number of observations~$E_{ij}$ of a particular edge and the posterior probability~$Q_{ij}$.  As the figure shows, an edge observed only zero or one times implies a low~$Q_{ij}$ (less than~0.1), so a single observation is probably a false alarm.  But two or more observations of the same edge result in a much larger $Q_{ij}$ (greater than~0.9), indicating a strong inference that the edge exists in the ground truth.  The sharp transition between low and high values of $Q_{ij}$ means that it is possible to infer the presence or absence of edges with good reliability despite the high error rate in the data.

For our second example we study a more traditional friendship network, taken from the National Longitudinal Study of Adolescent Health (the ``Add\-Health'' study)~\cite{Resnick97}.  This study compiled networks of friendships between students at a number of US high schools by asking participants to name their friends.  Again the data are both richer and poorer than a simple network measurement.  They are richer in the sense that we have two measurements of each friendship, from the point of view of each of the two participants, but poorer in the sense that those measurements can (and often do) disagree, indicating that respondents are not reliable in the reports they give or that they are employing different standards for what constitutes a friendship.

We represent this situation in our data model by giving each participant~$i$ their own individual true- and false-positive rates~$\alpha_i$ and~$\beta_i$.  Once again one can derive closed-form expressions for these parameters and for the posterior probabilities~$Q_{ij}$ of edges in the ground-truth network---see the Methods and Supplementary Materials.  The analysis can be applied to any of the schools in the Add\-Health study; we use one of the smaller schools as our example, solely because it allows us to make a clearer picture of the resulting network.

Again the EM algorithm converges quickly and reliably, giving a network-average true-positive rate $\bar\alpha=0.6083$, false-positive rate $\bar\beta=0.0096$, and prior edge probability $\rho=0.0235$.  These values indicate that non-existent friendships are rarely falsely reported as existing (low average~$\beta_i$), although, once again, arguably the more interesting quantity is the false discovery rate, the probability of a friendship that \emph{is} reported being false.  This probability, which is equal to $(1-\rho)\beta_i/[\rho\alpha_i+(1-\rho)\beta_i]$, is significantly larger, having a network average of~$0.3309$.  In other words, about one in three reported friendships doesn't really exist.  There is also a relatively high rate of failure to report friendships that do exist (many of the $\alpha_i$~are significantly less than~1).  The latter is perhaps less surprising given the design of the study: students were limited to naming at most ten friends, so those with more than ten would be obliged to omit some.

Figure~\ref{fig:results}c shows the inferred network of friendships, with edge widths again indicating the probability~$Q_{ij}$ that an edge exists, and node sizes now varying to indicate how reliable the nodes are, in terms of the fraction of reported friendships that actually exist (which is equal to one minus the false discovery rate, also called the precision).  Reports made by nodes depicted with large diameter are reliable, those made by smaller nodes are not.  Armed with these results, one can now calculate a multitude of further results, including any function of network structure.

These are just two examples of possible applications.  The particular data models applied here are quite general and could be applied to many other networks, but there are also other models one could use.  A range of further possibilities are discussed in the Supplementary Materials.

\subsection*{Methods}
In the first example given in the text, edge observations are independent (Bernoulli) random variables, conditioned on the ground truth~$A_{ij}$ for the appropriate node pair~$i,j$, with prior probability~$\rho$, true-positive rate~$\alpha$, and false-positive rate~$\beta$.  Suppose that for each node pair $i,j$ we make~$N_{ij}$ measurements and observe an edge to be present in~$E_{ij}$ of those measurements.  Then, under this independent edge model,
\begin{align}
P(\mbox{data}|\mat{A},\theta) &= \prod_{i<j} \bigl[ \alpha^{E_{ij}}
  (1-\alpha)^{N_{ij}-E_{ij}} \bigr]^{A_{ij}} \nonumber\\
  &\hspace{3em}{}\times \bigl[ \beta^{E_{ij}} (1-\beta)^{N_{ij}-E_{ij}}
   \bigr]^{1-A_{ij}}.
\label{eq:likelihood}
\end{align}
Given the prior probability~$\rho$ on the individual edges, the prior probability on the entire network is $P(\mat{A}|\rho) = \prod_{i<j} \rho^{A_{ij}} (1-\rho)^{1-A_{ij}}$.  We also assume that the prior probability distributions on $\alpha$, $\beta$, and~$\rho$ are all uniform in the interval~$[0,1]$.  Combining Eqs.~\eqref{eq:bayes} and~\eqref{eq:likelihood} we then have
\begin{align}
P(\mat{A},\theta|\mbox{data}) &= {1\over P(\mbox{data})}
  \prod_{i<j} \bigl[ \rho \alpha^{E_{ij}}
  (1-\alpha)^{N_{ij}-E_{ij}} \bigr]^{A_{ij}} \nonumber\\
  &\hspace{2em}{}\times \bigl[ (1-\rho) \beta^{E_{ij}} (1-\beta)^{N_{ij}-E_{ij}} \bigr]^{1-A_{ij}}.
\label{eq:model1like}
\end{align}

Taking the log, substituting into Eq.~\eqref{eq:mstep}, and differentiating with respect to~$\alpha$, we get
\begin{equation}
\sum_{\mat{A}} q(\mat{A}) \sum_{i<j} A_{ij} \biggl( {E_{ij}\over\alpha}
  - {N_{ij}-E_{ij}\over1-\alpha} \biggr) = 0.
\label{eq:alphadiff}
\end{equation}
Defining the posterior probability of an edge between $i$ and~$j$ by $Q_{ij} = P(A_{ij}=1|\mbox{data},\theta) = \sum_{\mat{A}} q(\mat{A}) A_{ij}$ and rearranging Eq.~\eqref{eq:alphadiff}, we then get
\begin{equation}
\alpha = {\sum_{i<j} E_{ij} Q_{ij}\over\sum_{i<j} N_{ij} Q_{ij}}.
\label{eq:methodsalpha}
\end{equation}
Similarly differentiating with respect to $\beta$ and $\rho$ we arrive at
\begin{equation}
\beta = {\sum_{i<j} E_{ij} (1-Q_{ij})\over\sum_{i<j} N_{ij} (1-Q_{ij})},
\qquad
\rho = {1\over{n\choose2}} \sum_{i<j} Q_{ij}.
\label{eq:methodsbeta}
\end{equation}
For the ``reality mining'' data set the values of the $N_{ij}$ are all equal to the same number~$N$, in which case Eqs.~\eqref{eq:methodsalpha} and~\eqref{eq:methodsbeta} reduce to Eqs.~\eqref{eq:params1} and~\eqref{eq:params2}.

\begin{widetext}
To calculate~$q(\mat{A})$ we substitute~\eqref{eq:model1like} into Eq.~\eqref{eq:estep} to get
\begin{align}
q(\mat{A}) &= {\prod_{i<j} \bigl[ \rho \alpha^{E_{ij}}
  (1-\alpha)^{N_{ij}-E_{ij}} \bigr]^{A_{ij}}
  \bigl[ (1-\rho) \beta^{E_{ij}} (1-\beta)^{N_{ij}-E_{ij}} \bigr]^{1-A_{ij}}\over
  \sum_{\mat{A}} \prod_{i<j} \bigl[ \rho \alpha^{E_{ij}}
  (1-\alpha)^{N_{ij}-E_{ij}} \bigr]^{A_{ij}}
  \bigl[ (1-\rho) \beta^{E_{ij}} (1-\beta)^{N_{ij}-E_{ij}} \bigr]^{1-A_{ij}}}
  \nonumber\\
  &= \prod_{i<j} {\bigl[ \rho \alpha^{E_{ij}}
  (1-\alpha)^{N_{ij}-E_{ij}} \bigr]^{A_{ij}}
  \bigl[ (1-\rho) \beta^{E_{ij}} (1-\beta)^{N_{ij}-E_{ij}} \bigr]^{1-A_{ij}}\over
  \sum_{A_{ij}=0,1} \bigl[ \rho \alpha^{E_{ij}}
  (1-\alpha)^{N_{ij}-E_{ij}} \bigr]^{A_{ij}}
  \bigl[ (1-\rho) \beta^{E_{ij}} (1-\beta)^{N_{ij}-E_{ij}} \bigr]^{1-A_{ij}}}
  \nonumber\\
  & = \prod_{i<j} Q_{ij}^{A_{ij}} (1-Q_{ij})^{1-A_{ij}},
\end{align}
where
\begin{equation}
Q_{ij} = {\rho \alpha^{E_{ij}} (1-\alpha)^{N_{ij}-E_{ij}}\over
          \rho \alpha^{E_{ij}} (1-\alpha)^{N_{ij}-E_{ij}} +
          (1-\rho) \beta^{E_{ij}} (1-\beta)^{N_{ij}-E_{ij}}}.
\end{equation}
Note that if we make no measurements for a pair of nodes~$i,j$ so that $N_{ij}=E_{ij}=0$ (``missing data''), this expression correctly gives~$Q_{ij}$ equal to the prior edge probability~$\rho$.

In the second model used in the text, measurements of edges come from unilateral statements made by individual nodes, which may vary in their reliability.  We define separate true- and false-positive rates~$\alpha_i,\beta_i$ for statements made by each node~$i$, then, by a derivation similar to the one above, we can show that
\begin{equation}
\alpha_i = {\sum_j E_{ij} Q_{ij}\over \sum_j N_{ij} Q_{ij}},
  \qquad
\beta_i = {\sum_j E_{ij} (1-Q_{ij})\over\sum_j N_{ij}(1-Q_{ij})}, 
  \qquad
\rho = {1\over{n\choose2}} \sum_{i<j} Q_{ij},
\label{eq:model2}
\end{equation}
where the posterior edge probability~$Q_{ij}$ is given by
\begin{equation}
Q_{ij} = {\rho \alpha_i^{E_{ij}} (1-\alpha_i)^{N_{ij}-E_{ij}}
          \alpha_j^{E_{ji}} (1-\alpha_j)^{N_{ji}-E_{ji}} \over
          \rho \alpha_i^{E_{ij}} (1-\alpha_i)^{N_{ij}-E_{ij}}
          \alpha_j^{E_{ji}} (1-\alpha_j)^{N_{ji}-E_{ji}} +
          (1-\rho) \beta_i^{E_{ij}} (1-\beta_i)^{N_{ij}-E_{ij}}
          \beta_j^{E_{ji}} (1-\beta_j)^{N_{ji}-E_{ji}}}.
\end{equation}
A full derivation is given in the Supplementary Materials.
\end{widetext}

\begin{acknowledgments} The author thanks Elizabeth Bruch, George Cantwell, Travis Martin, Gesine Reinert, and Maria Riolo for useful comments.  This work was funded in part by the National Science Foundation under grants DMS--1407207 and DMS--1710848.
\end{acknowledgments}

{\bigskip\small This research uses data from Add Health, a program project directed by Kathleen Mullan Harris and designed by J. Richard Udry, Peter S. Bearman, and Kathleen Mullan Harris at the University of North Carolina at Chapel Hill, and funded by grant P01--HD31921 from the Eunice Kennedy Shriver National Institute of Child Health and Human Development, with cooperative funding from 23 other federal agencies and foundations.  Special acknowledgment is due Ronald R. Rindfuss and Barbara Entwisle for assistance in the original design.  Information on how to obtain the Add Health data files is available on the Add Health website (http://www.cpc.unc.edu/addhealth).  No direct support was received from grant P01--HD31921 for this analysis.}

\section*{Supplementary materials}
\renewcommand{\theequation}{S\arabic{equation}}
\setcounter{equation}{0}
\renewcommand{\thefigure}{S\arabic{figure}}
\setcounter{figure}{0}

\subsection{Additional results for the reality mining network}
Our EM algorithm works by finding the values of the model parameters that give the best fit of the data model to the observed data.  The method does not, however, guarantee that we will get a \emph{good} fit.  Even the best fit may still be a bad one if the model itself is not capable of capturing the form of the data.  As an analogy, imagine a set of data points on a graph that follow an intrinsically curved path across the page.  We can fit a straight line through such points, but even the best fit will not be a good fit.  There simply is no good fit of a straight line to curved data.

\begin{figure}
\begin{center}
\includegraphics[width=\columnwidth]{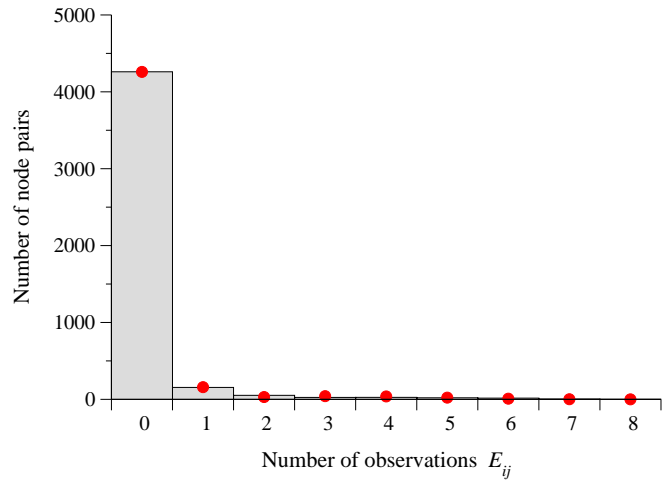}
\end{center}
\caption{Comparison of observations and fitted model.  The histogram shows the number of node pairs~$i,j$ with each possible value of~$E_{ij}$, the number of days on which the corresponding pair of individuals were observed in proximity.  The circles represent the predictions made by the model for the parameter values that give the best fit to the data.}
\label{fig:distribution}
\end{figure}

For the case of the reality mining data set of mobile phone proximities, a ``good fit'' to the data means one that captures accurately the numbers of proximity observations for pairs of individuals in the network.  Since the observations are assumed independent, only their number matters and not other features such as the specific days on which proximity is observed.  Figure~\ref{fig:distribution} shows a histogram of pairs of individuals in the network as a function of the number of days on which they are observed in proximity.  Because the network is sparse and a large majority of pairs never meet, most of the weight of the histogram is in the ``zero observations'' bin, although significant numbers of pairs fall in the other bins as well.  The circles show the values of the same quantities for the best-fit model---the one given by the parameter values in the paper.  As the figure shows, the fit is a good one, although there is some deviation between data and fit if one looks closely.

\begin{figure*}
\begin{center}
\null\hfill
\begin{minipage}{7cm}
\centering
\includegraphics[width=7cm]{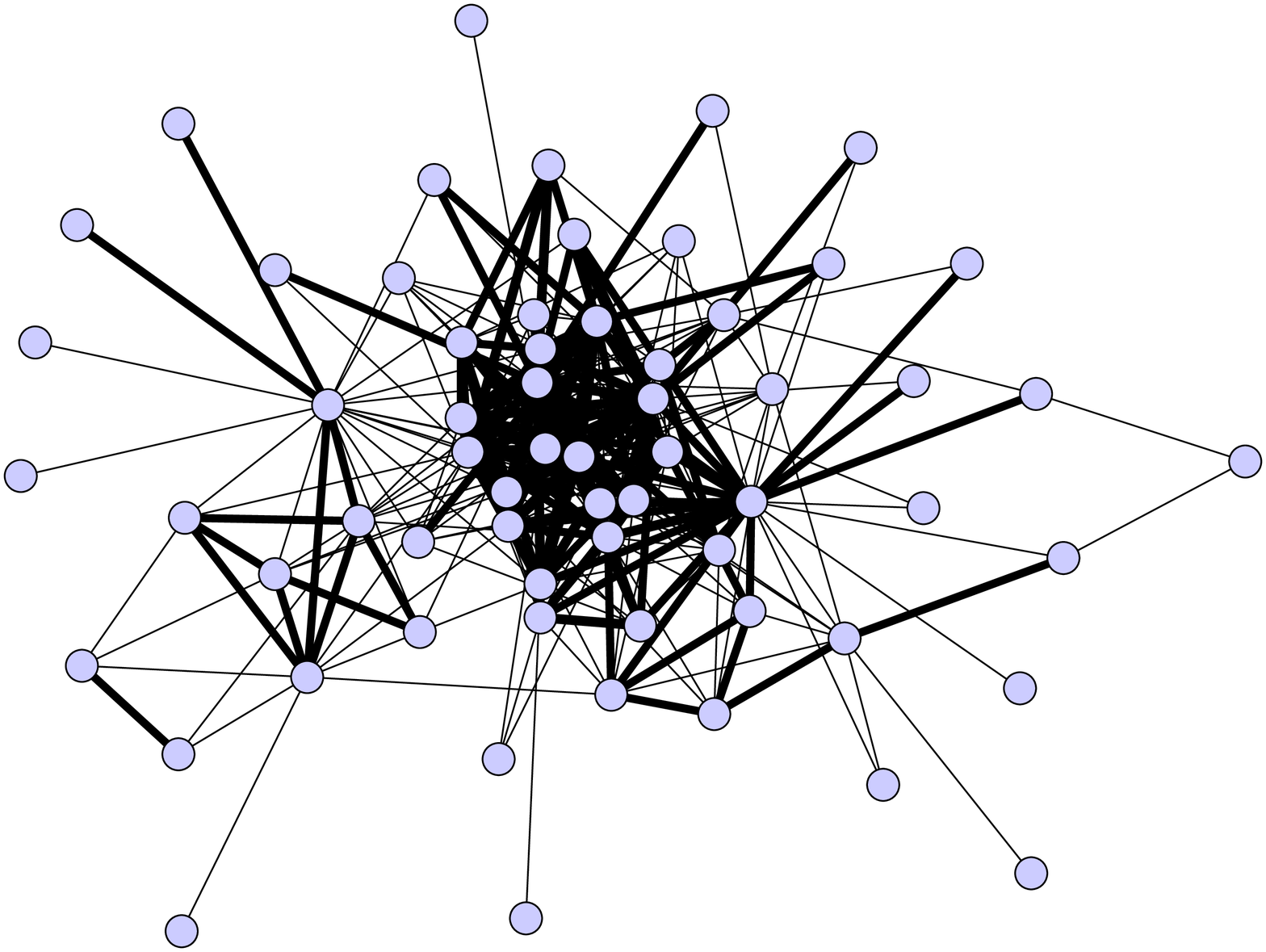}\\
{\small (a) Wednesdays}
\end{minipage}
\hfill
\begin{minipage}{7cm}
\centering
\includegraphics[width=7cm]{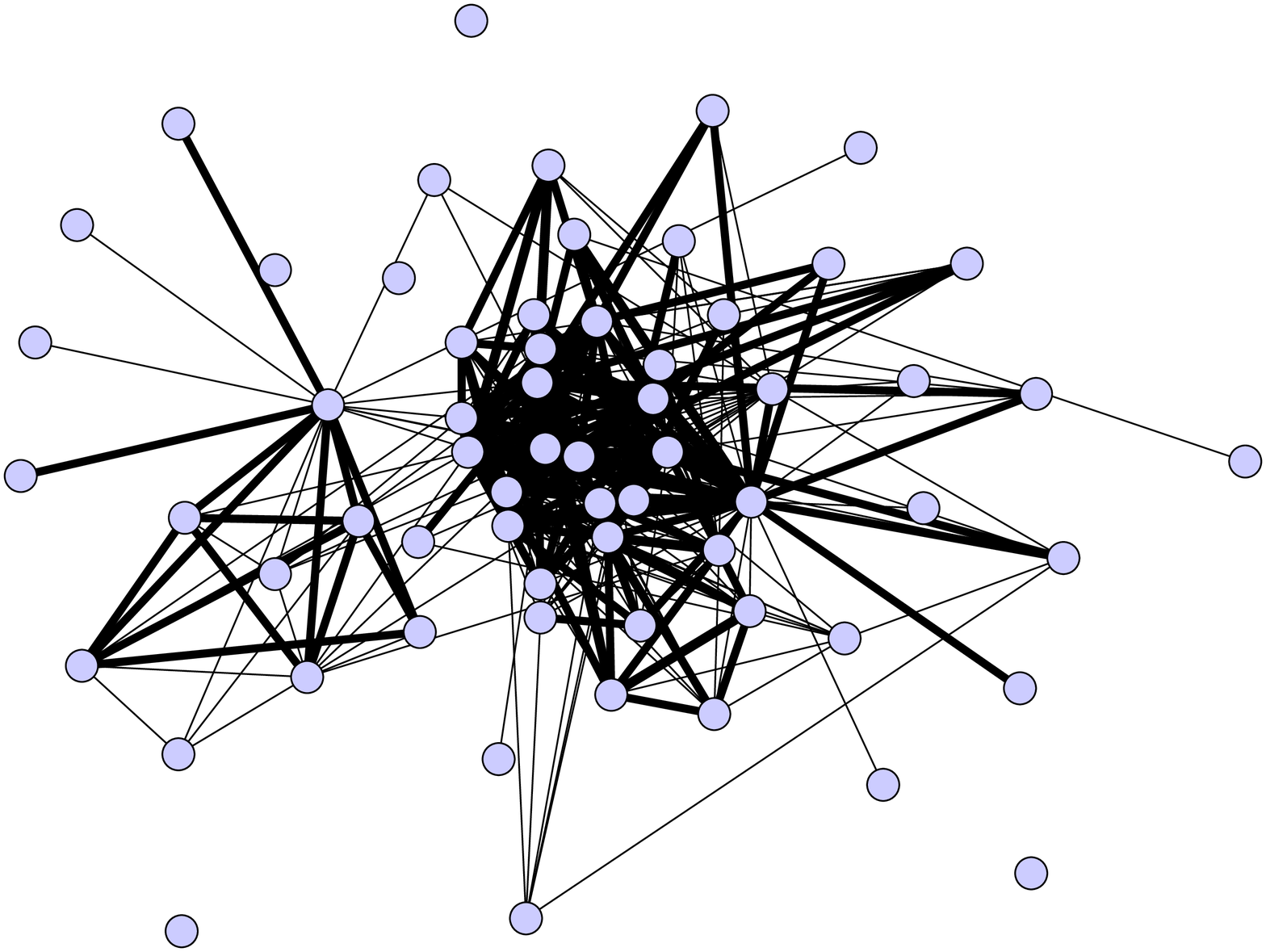}\\
{\small (b) Thursdays}
\end{minipage}
\hfill\null
\end{center}
\caption{(a)~The ``reality mining'' network reconstructed from proximity observations on eight consecutive Wednesdays, as described in the main text.  (This is the same network as in Fig.~\ref{fig:results}a, but redrawn for the purposes of this figure.)  (b)~The equivalent network reconstructed from observations on eight consecutive Thursdays during the same time period in March and April 2005.  We limit ourselves to the same set of nodes as in panel~(a), laid out in the same positions, to allow easy comparison between networks.}
\label{fig:wedthurs}
\end{figure*}

Another way to assess the quality of the results is to rerun the algorithm with an independent data set from the same source to see if we get a similar outcome.  A~nice feature of the reality mining study is that we have exactly such an independent data set available.  Recall that the results given in the paper are based on observations made on eight consecutive Wednesdays.  It is straightforward to perform the same analysis using data from a different day of the week.  Figure~\ref{fig:wedthurs} shows the network structure inferred from the Wednesday data (the same structure depicted in Fig.~\ref{fig:results}a) alongside the equivalent structure inferred from data for eight Thursdays over the same time period.  As the figure shows, the two networks are qualitatively similar, with a similar dense core and sparse periphery.  Some notable features, such as the tightly-connected satellite group of nodes to the left of center in the figure, are common to both networks.  But some individual details also differ from one network to the other---edges present in one are absent in the other and so forth.  This is natural: the whole point about error-prone data is that if we measure the same thing twice we do not expect to get exactly the same result.  Some variation between different measurements is expected, and indeed the extent of this variation can in principle be used to estimate the size of the experimental error.  In this case, however, we don't necessarily need to do this, since the model parameters---the true- and false-positive rates---already give us an estimate of the size of our errors.

\subsection{Additional results for the AddHealth network}
In the case of the Add\-Health friendship network studied in the paper, we considered the situation in which data obtained from different nodes may carry different degrees of reliability: in a social network of self-reported interactions such as this, some individuals may be more accurate in the information they give than others.  Let $E_{ij}$ be the number of times node~$i$ identifies node~$j$ as a friend.  (Normally this number will be zero or one, but we allow arbitrary numbers of identifications for the sake of generality.)  In effect, $E_{ij}$~constitutes a directed network, and self-reported friendship networks are sometimes depicted as being directed.
The underlying ground-truth network, however, is considered undirected.  Only our observations of it are directed.

A node whose identifications agree, wholly or largely, with those of their friends, is probably a reliable observer; one whose identifications disagree probably is not.  We do not have to impose these assumptions on our calculation, however.  They will automatically be reflected in the solution found by the EM algorithm, if they are a good description of the data.

\begin{widetext}
One can encapsulate this type of behavior in a model in which each node~$i$ has its own true-positive rate~$\alpha_i$ and false-positive rate~$\beta_i$.  Then the likelihood of a set of observations given a ground-truth network~$\mat{A}$ is
\begin{align}
P(\mbox{data}|\mat{A},\theta) &= \prod_{i<j} \bigl[ \alpha_i^{E_{ij}}
  (1-\alpha_i)^{N_{ij}-E_{ij}} \bigr]^{A_{ij}}
  \bigl[ \alpha_j^{E_{ji}} (1-\alpha_j)^{N_{ji}-E_{ji}} \bigr]^{A_{ji}} \nonumber\\
  &\hspace{4em}{}\times \bigl[ \beta_i^{E_{ij}}
  (1-\beta_i)^{N_{ij}-E_{ij}} \bigr]^{1-A_{ij}}
  \bigl[ \beta_j^{E_{ji}} (1-\beta_j)^{N_{ji}-E_{ji}} \bigr]^{1-A_{ji}},
\label{eq:model2like}
\end{align}
where $N_{ij}$ is the total number of observations of node~$j$ made by node~$i$.  Note that we explicitly include terms in $E_{ij}$ and $E_{ji}$ separately, since we want to incorporate observations made by all nodes.

Again assuming a prior probability of~$\rho$ on each ground-truth edge and uniform priors on the parameters, applying Eq.~\eqref{eq:bayes}, and taking logs, we arrive at the log-likelihood:
\begin{align}
\log P(\mat{A},\theta|\mbox{data}) &= \textstyle{\sum_{i<j}}
  \bigl[ A_{ij} E_{ij} \log \alpha_i + A_{ij} (N_{ij}-E_{ij}) \log(1-\alpha_i)
  + A_{ji} E_{ji} \log \alpha_j + A_{ji} (N_{ji}-E_{ji}) \log(1-\alpha_j)
  \nonumber\\
  &\hspace{6em}{}
  + (1-A_{ij}) E_{ij} \log \beta_i + (1-A_{ij}) (N_{ij}-E_{ij}) \log(1-\beta_i)
  \nonumber\\
  &\hspace{6em}{} + (1-A_{ji}) E_{ji} \log \beta_j
  + (1-A_{ji}) (N_{ji}-E_{ji}) \log(1-\beta_j)
  \nonumber\\
  &\hspace{6em}{} + A_{ij} \log\rho + (1-A_{ij}) \log(1-\rho) \bigr] - P(\mbox{data}).
\end{align}
Applying Eq.~\eqref{eq:mstep}, performing the derivatives, and rearranging, we then find the following best-fit values for the parameters:
\begin{equation}
\alpha_i = {\sum_j E_{ij} Q_{ij}\over \sum_j N_{ij} Q_{ij}},
  \qquad
\beta_i = {\sum_j E_{ij} (1-Q_{ij})\over\sum_j N_{ij}(1-Q_{ij})}, 
  \qquad
\rho = {1\over{n\choose2}} \sum_{i<j} Q_{ij},
\end{equation}
which recovers Eq.~\eqref{eq:model2} from the paper.  As before, $Q_{ij}$~is the posterior probability of an edge between $i$ and $j$, which can be calculated by the same method we used for our first model.  Noting that $A_{ij}=A_{ji}$ and combining Eqs.~\eqref{eq:bayes} and~\eqref{eq:model2like}, we write
\begin{align}
P(\mat{A},\theta|\mbox{data}) = {1\over P(\mbox{data})}
  &\prod_{i<j} \bigl[ \rho \alpha_i^{E_{ij}} (1-\alpha_i)^{N_{ij}-E_{ij}}
  \alpha_j^{E_{ji}} (1-\alpha_j)^{N_{ji}-E_{ji}} \bigr]^{A_{ij}} \nonumber\\
  &\qquad{}\times \bigl[ (1-\rho) \beta_i^{E_{ij}}
  (1-\beta_i)^{N_{ij}-E_{ij}} \beta_j^{E_{ji}} (1-\beta_j)^{N_{ji}-E_{ji}}
  \bigr]^{1-A_{ij}}.
\end{align}
Then the complete posterior distribution over ground-truth networks~$\mat{A}$ is
\begin{equation}
q(\mat{A}) = P(\mat{A}|\mbox{data},\theta)
  = {P(\mat{A},\theta|\mbox{data})\over
       \sum_{\mat{A}} P(\mat{A},\theta|\mbox{data})}
  = \prod_{i<j} Q_{ij}^{A_{ij}} (1-Q_{ij})^{1-A_{ij}},
\end{equation}
where
\begin{equation}
Q_{ij} = {\rho \alpha_i^{E_{ij}} (1-\alpha_i)^{N_{ij}-E_{ij}}
          \alpha_j^{E_{ji}} (1-\alpha_j)^{N_{ji}-E_{ji}} \over
          \rho \alpha_i^{E_{ij}} (1-\alpha_i)^{N_{ij}-E_{ij}}
          \alpha_j^{E_{ji}} (1-\alpha_j)^{N_{ji}-E_{ji}} +
          (1-\rho) \beta_i^{E_{ij}} (1-\beta_i)^{N_{ij}-E_{ij}}
          \beta_j^{E_{ji}} (1-\beta_j)^{N_{ji}-E_{ji}}}.
\end{equation}

Note that this expression is explicitly symmetric with respect to the indices~$i$ and~$j$, as it should be, since $Q_{ij}=Q_{ji}$ by definition.
\end{widetext}

This calculation returns not only an estimate of the ground-truth network but also an estimate of the trustworthiness of each of the nodes, parametrized by their true-positive and false-positive rates, which tell us both how often a node truthfully reports an edge that does exist and how often it falsely reports the existence of an edge that does not.  Note that even in the (common) case where each edge is observed at most once, so that $E_{ij}$ can take only the values zero and one, the parameters~$\alpha_i$ and~$\beta_i$ and the posterior probabilities~$Q_{ij}$ can take a wide range of values, by contrast with the case of the reality mining network, where there are only as many possible values of~$Q_{ij}$ as there are values of~$E_{ij}$ (see Fig.~\ref{fig:results}b).  For instance, even if both of nodes $i$ and $j$ report the existence of an edge between them ($E_{ij}=E_{ji}=1$), if neither node is considered reliable then the algorithm may say that the probability~$Q_{ij}$ of the edge actually existing is quite low.  If either of them is considered reliable, on the other hand, then~$Q_{ij}$ will be larger.  And if one is unreliable and claims an edge, while the other is reliable but does not, then $Q_{ij}$ will be particularly small.

\subsection{Other data models}
We have given two examples of possible data models.  There are however many others that could be used within the inference framework described, depending on the specific data available and the questions one wants to answer.

\paragraph*{Edge strengths or weights:} A simple variation on the model we used for the reality mining data set is one in which the underlying network can have edges with more than one strength.  A crude representation of the interactions of people in a social network, for instance, might record pairs of individuals only as ``acquainted'' or ``not acquainted.''  But a more nuanced representation might divide them into ``not acquainted,'' ``casual acquaintances,'' and ``well acquainted,'' and the frequency with which people meet might well differ between these classes: casual acquaintances might be more likely to meet than people who don't know each other at all, but less likely than people who are close friends.

Such a situation could be represented using a weighted adjacency matrix~$\mat{A}$ in which each element now has three possible values $0$, 1, and~2, with corresponding prior probabilities~$\rho_0$, $\rho_1$, and~$\rho_2$ such that $\rho_0+\rho_1+\rho_2=1$.  At the same time our two parameters~$\alpha$ and~$\beta$ would now become three---say $\alpha_0$, $\alpha_1$, and~$\alpha_2$---representing the probability of observing an edge in each of the three states.  With all other variables defined as before, the log-likelihood would then take the form
\begin{align}
\log P(\mat{A},\theta|\mbox{data}) &= \nonumber\\
  &\hspace{-5em} \textstyle{\sum_{i<j}} \bigl\lbrace
   \mathbbm{1}_{A_{ij}=0} \bigl[ E_{ij} \log\alpha_0 + (N_{ij}-E_{ij}) \log(1-\alpha_0) \bigr] \nonumber\\
  &\hspace{-5em}{} + \mathbbm{1}_{A_{ij}=1} \bigl[ E_{ij} \log\alpha_1
     + (N_{ij}-E_{ij}) \log(1-\alpha_1) \bigr] \nonumber\\
  &\hspace{-5em}{} + \mathbbm{1}_{A_{ij}=2} \bigl[ E_{ij} \log\alpha_2
     + (N_{ij}-E_{ij}) \log(1-\alpha_2) \bigr] \nonumber\\
  &\hspace{-5em}{} + \mathbbm{1}_{A_{ij}=0} \log \rho_0
    + \mathbbm{1}_{A_{ij}=1} \log \rho_1 + \mathbbm{1}_{A_{ij}=2} \log \rho_2
    \bigr\rbrace \nonumber\\
  &\hspace{-5em}{} - \log P(\mbox{data}),
\end{align}
where $\mathbbm{1}$ is the indicator function.  Then, applying Eq.~\eqref{eq:mstep} in the paper, we get
\begin{equation}
\alpha_w = {\sum_{i<j} E_{ij} Q_{ij}^{(w)}\over\sum_{i<j} N_{ij} Q_{ij}^{(w)}}, 
\qquad
\rho_w = {1\over{n\choose2}} \sum_{i<j} Q_{ij}^{(w)},
\label{eq:levels1}
\end{equation}
for $w=0, 1, 2$, where $Q_{ij}^{(w)} = \sum_{\mat{A}} \mathbbm{1}_{A_{ij}=w}\,q(\mat{A})$ is the posterior probability that $A_{ij}=w$, which is given by
\begin{equation}
Q_{ij}^{(w)} = {\rho_w \alpha_w^{E_{ij}} (1-\alpha_w)^{N_{ij}-E_{ij}}\over
          \sum_{w'} \rho_{w'} \alpha_{w'}^{E_{ij}} (1-\alpha_{w'})^{N_{ij}-E_{ij}}}.
\label{eq:levels3}
\end{equation}

This approach can easily be extended to any number of levels or strengths of connection between node pairs.  Equations~\eqref{eq:levels1} and~\eqref{eq:levels3} carry over unchanged.  Interesting questions arise about how we decide the ideal number of levels to include in the calculation (if we don't know \textit{a~priori}), which can be addressed using generalizations of standard model selection methods.  For instance, one could perform a $\chi^2$ test on the distribution of values of~$E_{ij}$, choosing the minimum number of levels for which the model is not rejected by the test to some predetermined degree of significance.

Note that within this framework the levels of the edges are not ordered: there is nothing in the mathematical formulation that stipulates that level~2 is ``stronger'' in any sense than level~1.  In practice this means that the parameter values returned by the EM algorithm may be permuted from the canonical order---all permutations give equally good fits to the data.  If we want the higher levels to correspond to stronger edges in the sense of greater values of~$\alpha_w$, then we may need to manually permute the levels after the algorithm completes its work.

Indeed, the levels need not correspond to strengths of edges at all.  Since the labels~$w$ on the levels are arbitrary and unordered, they could simply refer to different types of edges that happen to have different probabilities~$\alpha_w$ of observation.  In the case of social networks and proximity data, for instance, a network might include people who know each other through work, though family ties, and just socially, and one might find, for instance, that proximity is observed often for family members, a medium amount for coworkers, and more infrequently for other social ties.  In this case the EM algorithm, combined with the model described above, should be able to distinguish between these different types of connection, although it would not be able to tell us what the types actually represented---that would require additional interpretation by the experimenter, and possibly additional data as well.

\paragraph*{Multimodal data:} Another interesting possibility is that of ``multimodal'' network data, by which we mean data that quantify the structure of a network in several different ways, such as a social network probed using traditional interviews or questionnaires, and then probed again using data from an online social networking site.  An example is the Copenhagen Networks Study~\cite{Stopczynski14}, in which the interactions of a thousand individuals in Copenhagen were cataloged using measurements of face-to-face meetings, electronic communications, and online social networks.

Suppose that we have data that measures, directly or indirectly, a specific network~$\mat{A}$ in several different ways or modes, which we label by integers $m=1,2,3\ldots$\quad The existence, or non-existence, of an (undirected unweighted) edge between node pair~$i,j$ is measured~$N_{ij}^{(m)}$ times in mode~$m$.  (The most likely values are $N_{ij}^{(m)}=1$---the pair was observed once, the usual situation in most network studies---or $N_{ij}^{(m)}=0$---the case of ``missing data,'' where we have no information about a particular pair.  For the sake of generality, however, we once again allow the possibility of higher values.)  Generalizing our earlier models, we also define $E_{ij}^{(m)}$ to be the number of times an edge is actually observed between nodes~$i,j$ in mode~$m$, and we assume the measurements to be independent, both for different modes and for different nodes, conditioned on the underlying ground truth~$A_{ij}$.  But we allow for the (likely) situation in which measurements in different modes have different levels of accuracy, meaning that there are different true- and false-positive rates for each mode~$m$, which we will denote~$\alpha_m$ and~$\beta_m$.

\begin{widetext}
The log-likelihood for this model is given by
\begin{align}
\log P(\mat{A},\theta|\mbox{data}) &= \textstyle{\sum_{i<j}} \bigl\lbrace
  \textstyle{\sum_m} \bigl[ A_{ij} E_{ij}^{(m)} \log\alpha_m
     + A_{ij} (N_{ij}^{(m)}-E_{ij}^{(m)}) \log(1-\alpha_m) \nonumber\\
  & \hspace{5em}{} + (1-A_{ij}) E_{ij}^{(m)} \log\beta_m
   + (1-A_{ij})(N_{ij}^{(m)}-E_{ij}^{(m)}) \log(1-\beta_m) \bigr] \nonumber\\
  & \hspace{5em}{} + A_{ij} \log\rho + (1-A_{ij}) \log(1-\rho) \bigr\rbrace
     - \log P(\mbox{data}),
\end{align}
where $\rho$ is once again the prior probability of an edge.  Substituting this form into Eq.~\eqref{eq:mstep} and performing the derivatives, we get
\begin{equation}
\alpha_m = {\sum_{i<j} E_{ij}^{(m)} Q_{ij}\over\sum_{i<j} N_{ij}^{(m)} Q_{ij}},
\qquad
\beta_m = {\sum_{i<j} E_{ij}^{(m)} (1-Q_{ij})\over\sum_{i<j}
  N_{ij}^{(m)} (1-Q_{ij})},
\qquad
\rho = {1\over{n\choose2}} \sum_{i<j} Q_{ij},
\end{equation}
where $Q_{ij} = \sum_{\mat{A}} q(\mat{A}) A_{ij}$ is once again the posterior probability of an edge between nodes $i$ and~$j$.  Following the same line of argument as in the Methods, we find that
\begin{equation}
Q_{ij} = {\rho \prod_m \alpha_m^{E_{ij}^{(m)}}
   (1-\alpha_m)^{N_{ij}^{(m)}-E_{ij}^{(m)}}\over
   \rho \prod_m \alpha_m^{E_{ij}^{(m)}} (1-\alpha_m)^{N_{ij}^{(m)}-E_{ij}^{(m)}} +
   (1-\rho) \prod_m \beta_m^{E_{ij}^{(m)}} (1-\beta_m)^{N_{ij}^{(m)}-E_{ij}^{(m)}}}.
\label{eq:modesqij}
\end{equation}
\end{widetext}

To understand how the different modes are weighted by the algorithm, it is helpful to consider the odds ratio for an edge between nodes $i$ and~$j$:
\begin{equation}
{Q_{ij}\over1-Q_{ij}}
  = {\rho\over1-\rho} \prod_m
    \biggl( {\alpha_m\over\beta_m} \biggr)^{E_{ij}^{(m)}}
    \biggl( {1-\alpha_m\over1-\beta_m} \biggr)^{N_{ij}^{(m)}-E_{ij}^{(m)}}.
\end{equation}
Note how, in modes~$m$ for which $\alpha_m$ is large and $\beta_m$ is small, the $E_{ij}^{(m)}$ observed edges contribute a large increase to the odds ratio (first term in parentheses) and the $N_{ij}^{(m)}-E_{ij}^{(m)}$ non-edges contribute a large decrease (second term).  These modes are precisely the reliable ones---those with high true-positive rates and low false-positive rates---and hence it is appropriate that they contribute strongly to our inference of the network structure.

\subsection{Computation of network properties}
\label{sec:properties}
The primary output of our EM algorithms is the posterior probability distribution~$q(\mat{A})$ over possible ground-truth networks.  Given this distribution, one can in principle calculate the expected value or distribution of any other quantity that depends on network structure, such as degree distributions, clustering coefficients, correlation measures, spectral properties, and so forth.  If we have some metric~$X(\mat{A})$ whose value is a function of the network structure~$\mat{A}$, then its expected value, given the observed data, is
\begin{equation}
\mu_X = \sum_{\mat{A}} q(\mat{A}) X(\mat{A}),
\end{equation}
and the variance about that expectation is
\begin{equation}
\sigma_X^2 = \sum_{\mat{A}} q(\mat{A}) [X(\mat{A})-\mu]^2.
\end{equation}
These expressions are primarily of use for quantities whose distribution is approximately normal.  In other cases one can compute the complete probability distribution of~$X$ thus:
\begin{equation}
P(X=x|\mbox{data},\theta) = \sum_{\mat{A}} q(\mat{A}) \mathbbm{1}_{X(\mat{A})=x},
\end{equation}
where $\mathbbm{1}$ is the indicator function again.

In some cases is it possible to employ these expression directly.  Take for example, the calculation of the degree of a given node~$i$.  For the two data models employed in the paper, or any other model for which one can derive an explicit expression for the marginal probability~$Q_{ij}$ of an edge between two nodes, we can write the expected degree of node~$i$ as
\begin{equation}
d_i = \sum_{\mat{A}} q(\mat{A}) \sum_j A_{ij}
    = \sum_j \sum_{\mat{A}} q(\mat{A}) A_{ij}
    = \sum_j Q_{ij}.
\end{equation}

In other cases, particularly those in which the quantity of interest is a non\-local function of network structure, such as a correlation function or an eigenvalue, it may not be possible to perform the sum over networks~$\mat{A}$ in closed form, in which case one can estimate expectations, variances, or complete distributions using Monte Carlo sampling, whereby one draws a number of networks from the posterior distribution~$q(\mat{A})$, computes the quantity of interest on each of them, and then calculates the desired statistics.

In the particular case in which the posterior distribution factors into independent distributions over each edge---as in all of the models considered here---Monte Carlo sampling of networks is trivial.  One simply generates each edge independently with the appropriate probability~$Q_{ij}$, and there exist straightforward algorithms for doing this efficiently~\cite{REG17}.  In cases where the edges are not independent, one can generate networks using Markov chain importance sampling~\cite{NB99}, in which one repeatedly makes small changes~$\mat{A}\to\mat{A}'$ to the network, such as the addition or removal of a single edge, then accepts those changes with the standard Metropolis--Hastings acceptance probability
\begin{equation}
P_a = \biggl\lbrace\begin{array}{ll}
      q(\mat{A}')/q(\mat{A}) & \qquad\text{if $q(\mat{A}')<q(\mat{A})$,} \\
      1                      & \qquad\text{otherwise.}
      \end{array}
\end{equation}


\begin{thebibliography}{10}
\expandafter\ifx\csname url\endcsname\relax
  \def\url#1{\texttt{#1}}\fi
\expandafter\ifx\csname urlprefix\endcsname\relax\def\urlprefix{URL }\fi

\bibitem{Uetz00}
P.~Uetz~\etal, A comprehensive analysis of protein--protein interactions in saccharomyces cerevisiae. \textit{Nature} \textbf{403}, 623--627 (2000).

\bibitem{Ito01}
T.~Ito, T.~Chiba, R.~Ozawa, M.~Yoshida, M.~Hattori, and Y.~Sakaki, A
  comprehensive two-hybrid analysis to explore the yeast protein interactome.
  \textit{Proc. Natl. Acad. Sci. USA} \textbf{98}, 4569--4574 (2001).

\bibitem{Giot03}
L.~Giot, J.~S. Bader, C.~Brouwer, \textit{et~al.}, A protein interaction map of
  {D}rosophila melanogaster. \textit{Science} \textbf{302}, 1727--1736 (2003).

\bibitem{vonMering05}
C.~von Mering, L.~J. Jensen, B.~Snel, S.~D. Hooper, M.~Krupp, M.~Foglierini,
  N.~Jouffre, M.~A. Huynen, and P.~Bork, {STRING}: Known and predicted
  protein-protein associations, integrated and transferred across organisms.
  \textit{Nucleic Acids Research} \textbf{33}, D433--D437 (2005).

\bibitem{Krogan06}
N.~J. Krogan~\etal, Global landscape of protein complexes in the yeast {S}accharomyces cerevisiae. \textit{Nature} \textbf{440}, 637--643 (2006).

\bibitem{WPVS09}
S.~J. Wodak, S.~Pu, J.~Vlasblom, and B.~S\'eraphin, Challenges and rewards of
  interaction proteomics. \textit{Molecular \& Cellular Proteomics} \textbf{8},
  3--18 (2009).

\bibitem{RH61}
A.~Rapoport and W.~J. Horvath, A study of a large sociogram. \textit{Behavioral
  Science} \textbf{6}, 279--291 (1961).

\bibitem{Resnick97}
M.~D. Resnick~\etal, Protecting adolescents from harm: Findings from the
  {N}ational {L}ongitudinal {S}tudy on {A}dolescent {H}ealth. \textit{Journal
  of the American Medical Association} \textbf{278}, 823--832 (1997).

\bibitem{KB76}
P.~D. Killworth and H.~R. Bernard, Informant accuracy in social network data.
  \textit{Human Organization} \textbf{35}, 269--286 (1976).

\bibitem{BK77}
H.~R. Bernard and P.~D. Killworth, Informant accuracy in social network data
  {II}. \textit{Human Communications Research} \textbf{4}, 3--18 (1977).

\bibitem{Marsden90}
P.~V. Marsden, Network data and measurement. \textit{Annual Review of
  Sociology} \textbf{16}, 435--463 (1990).

\bibitem{Butts03}
C.~T. Butts, Network inference, error, and informant (in)accuracy: A {B}ayesian
  approach. \textit{Social Networks} \textbf{25}, 103--140 (2003).

\bibitem{DLR77}
A.~P. Dempster, N.~M. Laird, and D.~B. Rubin, Maximum likelihood from
  incomplete data via the {EM} algorithm. \textit{J. R. Statist. Soc. B}
  \textbf{39}, 185--197 (1977).

\bibitem{MK08}
G.~J. McLachlan and T.~Krishnan, \textit{The EM Algorithm and Extensions}.
  Wiley-Interscience, New York, 2nd edition (2008).

\bibitem{EP06}
N.~Eagle and A.~Pentland, Reality mining: Sensing complex social systems.
  \textit{Journal of Personal and Ubiquitous Computing} \textbf{10}, 255--268
  (2006).

\bibitem{Stopczynski14}
A.~Stopczynski, V.~Sekara, P.~Sapiezynski, A.~Cuttone, M.~M. Madsen, J.~E.
  Larsen, and S.~Lehmann, Measuring large-scale social networks with high
  resolution. \textit{PLOS One} \textbf{9}, e95978 (2014).

\bibitem{REG17}
A.~S. Ramani, N.~Eikmeier, and D.~F. Gleich, Coin-flipping, ball-dropping, and
  grass-hopping for generating random graphs from matrices of edge
  probabilities. Preprint arxiv:1709.03438 (2017).

\bibitem{NB99}
M.~E.~J. Newman and G.~T. Barkema, \textit{Monte Carlo Methods in Statistical
  Physics}. Oxford University Press, Oxford (1999).

\end{thebibliography}
\end{document}